\documentclass[twocolumn,prl,showpacs]{revtex4}
\usepackage{graphicx}
\usepackage{times}
\usepackage{amsmath,amssymb}

\newcommand{\eps}{\varepsilon}

\begin{document}

\title{Maximizing the N\'eel temperature of fermions in a simple-cubic optical lattice}

\author{C. J. M. Mathy}
\affiliation{Department of Physics, Princeton University,
Princeton, New Jersey 08544}
\author{David A. Huse}
\affiliation{Department of Physics, Princeton University,
Princeton, New Jersey 08544}

\date{\today}

\begin{abstract}
For a simple-cubic optical lattice with lattice spacing
$d$, occupied by two species of fermionic atoms of mass $m$ that
interact repulsively, we ask what conditions maximize the N\'eel
temperature $T_N$ in the Mott insulating phase at density one atom
per site, with equal numbers of the two species.  This maximum, $k_BT_N^{(\rm max)}\cong 0.15\,
\hbar^2/(md^2)$, occurs near the edge of the regime where the
system is well-approximated by the usual Hubbard model. The
correction to the Hubbard-model approximation that produces a ``direct'' ferromagnetic interaction between atoms in
nearest-neighbor Wannier orbitals is the leading term that limits how high $T_N$ can be made.
\end{abstract}

\pacs{}


\maketitle


One of the next notable milestones in the production of new
strongly-correlated many-body states with ultracold atoms is
expected to be the antiferromagnetic N\'eel phase of two hyperfine
species of fermionic atoms in an optical lattice \cite{duan,
bloch}. Important progress towards this goal includes the recent realization of the Mott insulating phase with fermions \cite{jordens}, and the
demonstration of controllable superexchange interactions in an
optical lattice, albeit with bosonic atoms \cite{trotsky}.  When
the optical lattice is sufficiently deep and the repulsive
$s$-wave interaction between the atoms is sufficiently weak, the
N\'eel temperature $T_N$ for the case of one atom per lattice site
can be estimated by modeling the system as a one-band Hubbard
model, and one can analyze the possibility of reaching this phase
by adiabatically ramping up the interactions and the optical
lattice \cite{duan,werner,dare,koetsier,snoek}. The most
accessible conditions for first producing this ordered phase in an
experiment will most likely be some compromise between the highest
$T_N$ and the highest entropy at the transition $S(T_N)$. If the
parameters of the system, namely the intensity $V_0$ of the
optical lattice and the $s$-wave scattering length $a_s$, can be
tuned in a perfectly adiabatic manner, then to access the N\'eel
phase only requires achieving sufficiently low entropy
\cite{werner,dare,koetsier}. But in the more likely event that
there is always some ``background'' heating, so things are not
perfectly adiabatic, the phase will be more accessible when it
occurs at higher absolute temperature. Thus in this paper we study
how the N\'eel temperature $T_N$ depends on the two tunable
parameters $V_0$ and $a_s$ as one leaves the region where the
standard Hubbard model is a good approximation to this system.
Away from the Hubbard regime, theoretical studies have suggested
that one may be able to access a wealth of phases governed by
quantum spin hamiltonians \cite{duan,duan2,duan3}.

According to quantum Monte Carlo simulations \cite{staudt} of the
simple-cubic fermionic Hubbard model, for a given nearest-neighbor
hopping matrix element $t$ the highest $k_BT_N\cong t/3$ occurs at
interaction $U\cong 8\,t$, while for a given $U$ the maximal
$k_BT_N\cong U/20$ occurs at $t\cong 0.15\,U$.  Thus to increase
$T_N$ one wants to move to larger $t$, which means a weaker
optical lattice (smaller $V_0$), and to larger $U$, which means
larger $a_s$. This necessarily moves the system away from the
regime where it is well-approximated by the usual Hubbard model.
The mapping from the real system to the Hubbard model uses the
single-atom Wannier orbitals as the basis states
\cite{kohn,duan,werner}. The standard one-band Hubbard model
includes only the lowest-energy Wannier orbital at each lattice
site and only the on-site interaction between two atoms of
different hyperfine states occupying the same Wannier orbital.

We find that it is the corrections due to including the
interactions between Wannier orbitals on nearest-neighbor lattice
sites that are the leading effects that stop and reverse the
increase of $T_N$ as $t$ and $U$ are increased by decreasing $V_0$
and increasing $a_s$. In particular, these interactions produce a
``direct'' ferromagnetic exchange interaction favoring neighboring
sites to be occupied by the same species. These ferromagnetic
interactions are of the opposite sign from the antiferromagnetic
superexchange interactions that cause the N\'eel ordering, and
thus they suppress $T_N$.  Within the approximations that we make
(discussed in detail below) the maximal $k_BT_N^{\rm
(max)}\cong 0.03 E_r$ occurs near $V_0\cong 3E_r$ and $a_s\cong
0.15\, d$, where $E_r=(\pi\hbar)^2/(2md^2)$ is the recoil energy
and $d$ is the lattice spacing. For example, $E_r\cong1.4\, \mu$K
for $^6$Li with $d=532$ nm, which puts the maximum N\'eel temperature near 40 nK, which seems to be well within the reach of current experimental cooling techniques.

This regime of large repulsive $a_s$ is attained by approaching a
Feshbach resonance from the repulsive side.  But the atoms must
scatter repulsively without ``falling'' in to the weakly-bound
molecular state.  Ref. \cite{jordens} studied the Mott insulator with $^{40}$K at $a_s\cong 0.08\, d$ and did not mention any problem with excessive molecule formation.  It is not clear whether this can be increased to the
$a_s\cong
0.15\, d$ that maximizes $T_N$ \cite{petrov}.  It is also not clear whether the
optical lattice increases or decreases molecule formation. 
The
lattice breaks momentum conservation, thus possibly opening up
channels for molecule formation, while in the Mott insulator the
atoms are kept apart in different wells of the optical
lattice, which, naively, reduces the opportunities for molecule
formation.  

The system we consider is made up of fermionic atoms in a simple-cubic
optical lattice with a single-atom potential of the standard form
\cite{bloch}:
\begin{equation}
V_1(x,y,z)=V_0(\sin^2\frac{\pi x}{d}+\sin^2\frac{\pi
y}{d}+\sin^2\frac{\pi z}{d}) ~.
\end{equation}
This is a separable potential, so the energy eigenstates of a
single atom in this potential can be chosen to be the product of
one-dimensional (1D) eigenstates along each 
direction.  We solve for these 1D bands and thus obtain the
properly normalized wavefunctions $w_n(x)$ of the maximally-localized 1D Wannier orbitals for each band $n$ \cite{kohn}. 
The 3D Wannier orbitals are then
$\phi_{n_xn_yn_z}(x,y,z)=w_{n_x}(x)w_{n_y}(y)w_{n_z}(z)$.

We will be focusing on the Mott insulating regime with density
exactly one atom per lattice site and low temperature, where the
atoms are primarily in the lowest band, $n_x=n_y=n_z=0$. The
nearest-neighbor hopping matrix element $t_0$ in this band
is a strongly decreasing function of $V_0$, the lattice strength, behaving as
$t_0\approx 4 \pi^{-1/2} E_r^{1/4} V^{3/4} e^{-2\sqrt{V/E_r}}$ for large $V_0$ \cite{werner}.
The wavefunction $w_0(x)$ of the lowest Wannier
orbital at a given lattice site is positive and has its maximum
amplitude at $x=0$, the center of the well of the optical lattice
at that site, while its amplitude is negative and of smaller magnitude 
in the nearest-neighbor wells of the
lattice.  The ratio of these amplitudes is one small parameter that is important in the approximations we use below.

We are interested in the antiferromagnetic Mott insulating phase
in the regime where the on-site repulsive interaction $U$ is
stronger than the hopping $t$.  We do not treat the limit
of weak interaction, where the system is a paramagnetic Fermi
liquid. We assume the atoms 
equally populate two
different hyperfine states; as is standard, we will call these two
states ``up'' and ``down'' and treat them as the two states of a
spin-1/2 degree of freedom.  The $s$-wave repulsive interaction is
only between atoms of opposite spin.  To lowest order in $(a_s/d)$, we
approximate this 2-atom interaction as the standard regularized contact potential\cite{Huang}
\begin{equation}
V_2(\vec r_{\uparrow}-\vec
r_{\downarrow})=\frac{4\pi\hbar^2a_s}{m}\delta(\vec
r_{\uparrow}-\vec r_{\downarrow}) \frac{\partial}{\partial r}r ~
\end{equation}
where r distance between the two particles.
The expectation value of this interaction energy for two atoms of
opposite spin occupying the same lowest Wannier orbital is our
first estimate of the strength of the on-site interaction
$U_0n_{i\uparrow}n_{i\downarrow}$ 
in the corresponding
one-band Hubbard model:
\begin{equation}
U_0=\frac{4\pi\hbar^2a_s}{m}[\int dxw_0^4(x)]^3 ~.
\end{equation}

In the Hubbard model, when adjacent sites $i$ and $j$ are each
singly-occupied by atoms with the same spin, then the hopping between those
two sites is Pauli-blocked.  When these adjacent sites are each
singly-occupied by opposite spins, then virtual hopping between
these sites, treated in second-order perturbation theory, allows
them to lower their energy and thus generates an antiferromagnetic
superexchange interaction $J_s(\vec S_i\cdot\vec S_j-\frac{1}{4})$
with $J_s=4t^2/U$.

\begin{figure}
\includegraphics[width=0.5\textwidth]{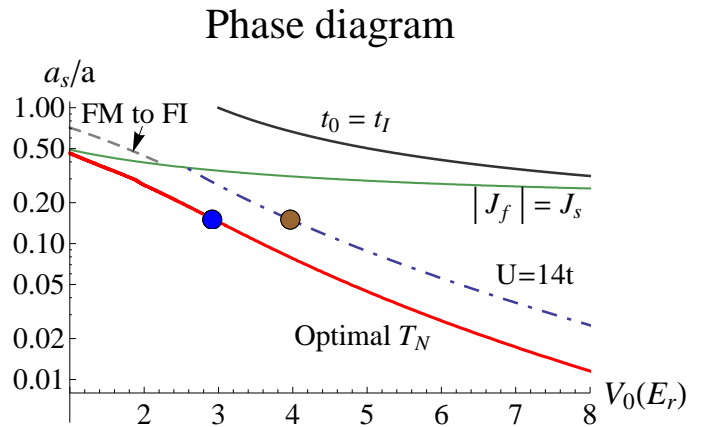}
\caption{Approximate phase diagram for filling one fermion per lattice site. 
The line marked $|J_f|=J_s$ is our approximation to the
ground-state phase boundary separating the antiferromagnetic phase
at smaller $a_s$ from the ferromagnetic phase at larger $a_s$. The
ferromagnetic phase is mostly a fully-polarized band insulator,
but there is a small sliver of polarized Fermi liquid at small
$V_0$ between the lines marked $|J_f|=J_s$ and ``FM to FI". The
line marked ``Optimal $T_N$" indicates where $T_N$ as a function
of $V_0$ is maximized for each given $a_s$.  The dot on that line
is our estimate of the parameters that produce the overall maximum
of $T_N/E_r$, and at that point $J_f \simeq -J_s/4$ (see text).  The $U=14t$ line is near where the entropy is maximized
at $T_N$ \cite{werner} and $T_N$ on this line is maximized at the dot. 
The $t_0=t_I$ line signals when the interaction correction to the hopping 
becomes strong.  There is presumably also a paramagnetic Fermi liquid ground state in the
lower left corner of this phase diagram, but our approximations are not well-suited
to estimating where this phase is.} 
\label{fig:PhaseDiag}
\end{figure}

The leading corrections to the Hubbard model approximation to this
system in the regime we are interested in are due to the
interactions between atoms of opposite spin occupying lowest
Wannier orbitals on nearest-neighbor sites $i$ and $j$.  There are
2 contributions: First, and apparently most important in limiting how large $T_N$ can be made, is the ``direct'' interaction \cite{trotsky}
\begin{equation}
U_{nn}=\frac{4\pi\hbar^2a_s}{m}[\int dxw_0^4(x)]^2\int
dyw_0^2(y)w_0^2(y+d)
\end{equation}
between atoms of opposite spin in adjacent orbitals.  This term is due to the overlap of the probability distributions of adjacent Wannier orbitals.  It
raises the energy of the N\'eel state.  It thus produces a direct
ferromagnetic exchange interaction $J_f(\vec S_i\cdot\vec
S_j-\frac{1}{4})$ with $J_f=-2U_{nn}<0$ that partially cancels the
antiferromagnetic superexchange $J_s$ that occurs in the Hubbard
model.  It is primarily this ferromagnetic interaction that stops
and reverses the increase in $T_N$ as one moves towards stronger
interaction and a weaker lattice while staying near the optimal
values of $U/t$. At the global maximum of $T_N$, indicated in Fig.
\ref{fig:PhaseDiag}, we find $J_f \simeq -J_s/4$.

For large enough $a_s$ this direct ferromagnetic
exchange is stronger than the superexchange and thus we have a
ground-state phase transition to a ferromagnetic phase, as
indicated in Fig. 1, and discussed more below.  [Very near this $|J_f|=J_s$ line, effects due
to weaker further-neighbor interaction might produce some
other magnetically-ordered phases.]  

The ``direct'' nearest-neighbor interaction (4) also reduces the
effective $U$ that enters in the superexchange interaction, so at
this level of approximation our simple-cubic Hubbard model has
interaction $U=U_0-6U_{nn}$, since it is the {\it change} of the
interaction energy due to moving the atom that enters in the
energy denominator in the superexchange process.

Also, the interaction generates an additional hopping term of the
same sign as $t_0$: \cite{werner}
\begin{equation}
t_I=-\frac{4\pi\hbar^2a_s}{m}[\int dxw_0^4(x)]^2\int
dyw_0^3(y)w_0(y+d)
\end{equation}
that is operative when the two sites are each singly-occupied by
opposite spins. The resulting effective hopping that enters in the
superexchange process at this level of approximation is thus
$t=t_0+t_I$.

Thus once we include these leading effects due to the
nearest-neighbor interaction, the effective Hamiltonian in the vicinity of the ground state of the half-filled Mott insulator has hopping $t=t_0+t_I$, an
effective on-site interaction $U=U_0-6U_{nn}$, and an additional
ferromagnetic nearest-neighbor exchange interaction $J_f=-2U_{nn}$
when both sites are singly-occupied.  To estimate the N\'eel
temperature of our system we propose the following approximation:
For the Hubbard model without $J_f$, we have estimates of its
N\'eel temperature $T_N^{(H)}(t,U)$ from quantum Monte Carlo
simulations \cite{staudt}.  This N\'eel ordering is due to the
antiferromagnetic superexchange interaction $J_s=4t^2/U$ between
neighboring singly-occupied sites. When we include $J_f<0$ this
reduces this magnetic interaction, and we will approximate the
resulting reduction of $T_N$ as being simply in proportion to the
reduction of the total nearest-neighbor exchange interaction:
\begin{equation}
T_N(V_0,a_s)\cong(1+\frac{J_f}{J_s})T_N^{(H)}(t,U) ~.
\end{equation}

\begin{figure}
\includegraphics[width=0.5\textwidth]{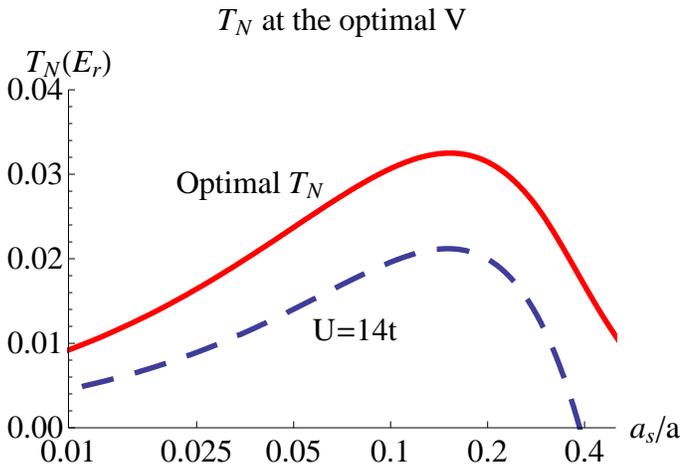}
\caption{Our estimates of the optimal N\'eel temperature, $T_N$,
as a function of $a_s/d$. For each value of $a_s$, $T_N$ is
maximized by varying the lattice depth $V_0$. We also plot $T_N$
at the line $U=14t$, which is near where the critical entropy is
maximized \cite{werner}.}
\end{figure}

In Fig. 1 we show the lattice strength $V_0/E_r$ that maximizes
this approximation to $T_N$ for each value of the interaction
$a_s/d$. The highest $T_N$ occurs near $a_s/d=0.15$, but the
system at this value of $a_s$ is may be too close to the
Feshbach resonance and thus not stable against formation of
molecules.  The highest achievable $T_N$ thus may be
somewhere along this line at a lower value of $a_s$ and thus a
stronger lattice $V_0$. We note that a recent experiment \cite{jordens} has studied $a_s/d \simeq 0.08$ for $^{40}K$, albeit at a temperature well above $T_N$, without noting any strong instability towards molecule formation.  We also show on Fig. 1 the line along
which $U=14t$, since this is near where the critical entropy
$S(T_N)$ is maximal \cite{werner}, so if the system can be
adjusted adiabatically this is where the N\'eel phase is most
accessible.

In Fig. 2 we show $k_BT_N/E_r$ as a function of
$a_s/d$ at the value of $V_0$ that maximizes our estimate of
$T_N$, as well as at the value of $V_0$ that gives $U=14t$ and
thus is near the maximum of $S(T_N)$.  Note that in Fig. 2 the horizontal scale for $a_s/d$ is logarithmic, so $T_N$ drops rather weakly as $a_s$ is reduced, meaning that the possible limitation in how large $a_s$ can be made will not ``cost'' a lot in terms of the resulting reduction of $T_N$.

The approximations we are making are clearly beginning to break
down in the vicinity of the parameter values that maximize $T_N$.  Thus, although we expect that these approximations give reasonably reliable rough estimates of the 
maximal values of $T_N$, there are many higher-order effects that we are ignoring that may alter these estimates by a little (our calculations suggest on the $\sim 10\%$ level).
At the maximum of $T_N$, $|J_f|$ is about 25\% of $J_s$.  The correction to $J_s$ due to $t_I$ is also of roughly this size, but its dependence on $a_s$ is much weaker, which is why $J_f$ is the important actor in causing the maximum in $T_N$.

\begin{figure}[htb]
\includegraphics[width=0.45\textwidth]{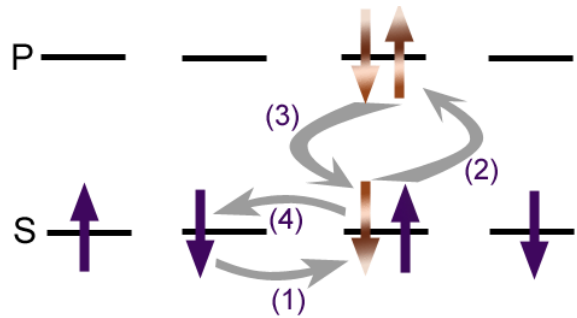}
\caption{The strongest higher-order process contributing to the
energy of the antiferromagnetic Mott insulator at the maxima of
$T_N$ shown in Fig. \ref{fig:PhaseDiag}.  It consists of (1) a
nearest-neighbor hop in the lowest (S) band, (2) an on-site ``pair
hopping'' of both fermions up to the next (P) band, (3) on-site
pair hopping back to the S band, and (4) a nearest-neighbor hop
back to the original configuration.  At both maxima of $T_N$, this
process corrects $J_s$ by about 10\%.}
\label{fig:VirtualHop6}
\end{figure}

\begin{table}[h]
\caption{The values of the various energies at the two $T_N$ maxima. The ``$J_s$ correction'' corresponds to the process detailed in fig. \ref{fig:VirtualHop6}.}
\begin{tabular}{c | c | c}
		&	Global maximum		& Maximum\\
		&	of $T_N$			& of $T_N$ at $U=14t$ \\ \hline
$\eps_0\, (E_r)$ &    4.2      &    5.5    \\
$U_0\,(E_r)$	&  0.9	&	1.3      \\
$t_0\,(E_r)$	&	0.11	&	0.07 \\
$t_I\, (E_r)$	&	0.02	&	0.02 \\
$J_s\, (E_r)$		&	0.08	&	0.03 \\
$J_f\, (E_r)$		&	-0.02	&	-0.008 \\
$J_s$ correction $(E_r)$  & 0.007	& 0.004 \\
\end{tabular}
\end{table}

The approximations we have used are those appropriate for the Mott insulator, and are based on the inequalities on energy scales $\eps_0>U>t$, where $\eps_0$ is the expectation value of the single-particle energy in a lowest Wannier orbital.
We have analyzed in perturbation theory many corrections beyond those included above.  We find that at 
the maximum of $T_N$ (both the global maximum and the maximum
along the $U=14t$ line) the strongest next correction is the
fourth-order process illustrated in Fig. \ref{fig:VirtualHop6}; it
alters $J_s$ by about 10\%. Since
our perturbatively-based approximations are breaking down near
this regime of interest where $T_N$ is maximized, it would be nice
to have a more systematic approach that can obtain more precise
and reliable estimates of the phase diagram in this regime.  For
example, quantum Monte Carlo simulations might be possible for
temperatures near $T_N$, although of course the famous fermionic
``minus signs'' may prevent this from being feasible in the near term.

The ferromagnetic phase of this model at strong repulsion is
mostly a band insulator, with a band gap between the
spin-polarized bands.  But in the weaker lattice regime there
should also be a partially-polarized Fermi liquid ground state
near the phase boundary to the N\'eel state.  The transition from
the fully-polarized band insulator to the partially-polarized
ferromagnet occurs when the spin-polarized bands overlap, so the
system can lower its energy by flipping spins.  A single spin flip
makes a hole and a doubly-occupied site that are each moving
freely within the fully-polarized background state.  At the level
of approximation we have used in this paper, the hole moves freely
with hopping $t_0$, so its lowest energy is $-6t_0$. The
doubly-occupied state costs interaction energy $U_0+6U_{nn}$ and
moves freely with effective hopping $t_2=t_0+2t_I$ because its
motion is the hopping of the flipped spin between sites that are
both occupied by unflipped spins.  The total energy of this
particle-hole pair can be negative when
$U_0+6U_{nn}<12(t_0+t_I)=6(t_0+t_2)$; this occurs below the line
indicated in Fig. 1 as ``FM to FI'' (ferromagnetic metal to
ferromagnetic insulator). We show this for completeness, although
these ferromagnetic phases at high $a_s$ are very likely to be inaccessible in
experiments with cold fermionic atoms in optical lattices.  Also, the present approximations are probably not very reliable in this regime of large $a_s/d$.

There is also a paramagnetic Fermi liquid phase at weak enough
lattice and at weak enough interaction, as well as possibly an
antiferromagnetic Fermi liquid near it.  These phases occur well
away from the regimes we have focussed on here, and the present
approximations are not well suited to estimating the location of
the corresponding phase transitions, so we leave that part of the
phase diagram as ``{\it terra incognita}'' for now.  The quantum
phase transition between the N\'eel state and the paramagnet
should occur in parameter regimes that are accessible to the
experiments, although it may not be possible to see its effects at
accessible temperatures, since $T_N$ decreases strongly as this
regime is approached.

{\it Conclusion:}  We have shown that to maximize the N\'eel
temperature one must leave the region of parameter space where the
Hubbard model approximation for fermionic atoms in an optical
lattice is well-controlled. We have found that the
nearest-neighbor direct ferromagnetic exchange is the most
important correction to the Hubbard model that limits the maximal $T_N$.
There are also higher-order corrections to the Hubbard model:
virtual hopping into higher bands and other higher-order processes. For
the parameters that maximize $T_N$, these higher-order terms are
smaller than the nearest-neighbor terms we include, although not
by a large margin of ``safety''.

The relative contribution of the higher-order corrections in the
vicinity of the optimal $T_N$ drops exponentially as one goes to
smaller interaction $a_s$ and thus a stronger optical lattice.
Thus our results are accurate in the large $V_0$ (strong lattice)
limit, and should qualitatively capture the phase diagram for
weaker lattices.  For quantitatively more accurate results in the
weak lattice regime, one needs to resort to more systematic
quantum calculations.  Of course this is a system of many
fermions, so it is not clear whether this weak lattice regime can
be accurately treated in some form of quantum Monte Carlo
simulations.


We thank Randy Hulet for many discussions, and Meera Parish for helpful suggestions.  This work was
supported under ARO Award W911NF-07-1-0464 with funds from the
DARPA OLE Program.


\end{document}